\documentclass[aip,jcp,reprint]{revtex4-2}

\usepackage[ngerman,main=english]{babel}

\usepackage[utf8]{inputenc}
\usepackage[T1]{fontenc}
\usepackage{lmodern}

\usepackage[inline]{enumitem}
\usepackage{graphicx}
\usepackage{hyperref}
\usepackage{orcidlink}
\usepackage{emptypage}

\usepackage{amsmath}
\usepackage{amssymb}
\usepackage{mathrsfs}
\usepackage{esint}
\usepackage{commath}
\usepackage{braket}
\usepackage{bbm}
\usepackage{booktabs}
\usepackage[section]{placeins}
\usepackage{appendix}
\usepackage{csquotes}
\usepackage{textgreek}
\usepackage{float}
\usepackage{multirow}
\usepackage{ragged2e}

\usepackage{tikz}
\usetikzlibrary{trees}

\usepackage{cleveref}

\crefname{equation}{Eq.}{Eqs.}
\crefname{figure}{Fig.}{Figs.}
\crefname{table}{Tab.}{Tabs.}
\crefname{section}{Sec.}{Secs.}
\crefname{subsection}{Sec.}{Secs.}
\crefname{appendix}{Appendix}{Appendices}

\newcommand{\mi}{\mathrm{i}}
\newcommand{\me}{e}

\newcommand{\ec}{\mathrm{e}}
\newcommand{\kB}{k_\mathrm{B}}

\newcommand{\diff}[1]{\mathrm{d}#1}

\newcommand{\dpar}[2][]{\frac{\partial #1}{\partial #2}}

\newcommand{\tr}{\operatorname{tr}}
\newcommand{\Hsys}{H_\mathrm{S}}
\newcommand{\Hbath}{H_\mathrm{B}}
\newcommand{\HSB}{H_\mathrm{SB}}
\newcommand{\rhosys}{\rho_\mathrm{S}}

\newcommand{\dblket}[1]{| #1 \rangle\!\rangle}
\newcommand{\dblbra}[1]{\langle\!\langle #1 |}
\newcommand{\dblbraket}[1]{\langle\!\langle #1 \rangle\!\rangle}

\newcommand{\meV}{\mathrm{meV}}

\makeatletter
\def\@email#1#2{
 \endgroup
 \patchcmd{\titleblock@produce}
  {\frontmatter@RRAPformat}
  {\frontmatter@RRAPformat{\produce@RRAP{*#1\href{mailto:#2}{#2}}}\frontmatter@RRAPformat}
  {}{}
}
\makeatother
\begin{document}
\title{Propagating the Hierarchical Equations of Motion using the Multi-Layer Multi-Configurational Time-Dependent Hartree method}
\author{Robin L. Grether\,\orcidlink{0009-0008-0172-1033}}
\email[Corresponding author: ]{robin.grether@physik.uni-freiburg.de}
\affiliation{Institute of Physics, Albert Ludwig University of Freiburg, Hermann-Herder-Str.~3, 79104 Freiburg im Breisgau, Germany}
\author{Samuel L. Rudge\,\orcidlink{0000-0002-2727-2191}}
\affiliation{Institute of Physics, Albert Ludwig University of Freiburg, Hermann-Herder-Str.~3, 79104 Freiburg im Breisgau, Germany}
\author{Martin Mäck\,\orcidlink{0009-0008-7221-5381}}
\affiliation{Institute of Physics, Albert Ludwig University of Freiburg, Hermann-Herder-Str.~3, 79104 Freiburg im Breisgau, Germany}
\author{Riley J. Preston\,\orcidlink{0009-0008-9858-4951}}
\affiliation{Institute of Physics, Albert Ludwig University of Freiburg, Hermann-Herder-Str.~3, 79104 Freiburg im Breisgau, Germany}
\author{Uwe Manthe\,\orcidlink{0000-0002-4845-9633}}
\affiliation{Theoretical Chemistry, Faculty of Chemistry, University of Bielefeld, Universitätsstr.~25, 33615 Bielefeld, Germany}
\author{Michael Thoss\,\orcidlink{0000-0002-3557-9855}}
\affiliation{Institute of Physics, Albert Ludwig University of Freiburg, Hermann-Herder-Str.~3, 79104 Freiburg im Breisgau, Germany}

\begin{abstract}
The hierarchical equations of motion (HEOM) are a powerful, numerically exact approach for simulating the time evolution of open quantum systems. Recently, several tensor-train- and tensor-network-based approaches have been suggested and realized to extend the range of applicability of the HEOM approach to larger and more realistic systems. In this paper, we build upon the existing twin-space formulation of the HEOM and present a novel approach to propagate the fermionic HEOM efficiently using the multi-configurational time-dependent Hartree method (MCTDH). We benchmark the resulting HEOM+MCTDH method with electron transport calculations for a Holstein-type nanojunction model. Furthermore, we demonstrate the capabilities of the HEOM+MCTDH method by applying it to a more complex vibronic coupling nanojunction model with strong molecule-electrode coupling for which fully quantum simulations have not been feasible previously.
\end{abstract}

\maketitle

\section{Introduction}\label{sec:introduction}

The theoretical description and computational simulation of open quantum systems remain a central challenge for understanding a variety of physical, chemical, and biological processes. In open quantum systems, a small quantum system of interest---for example a molecule or nanostructure---interacts with a surrounding environment characterized by its thermodynamic properties. Prominent examples include charge transport through nanostructures,\cite{reed_conductance_1997,nitzan_electron_2003,joachim_molecular_2005,breuer_theory_2007,hartle_multimode_2008,hartle_resonant_2011,hartle_decoherence_2013,wang_numerically_2013,hartle_formation_2014,hartle_transport_2015,schinabeck_hierarchical_2016,cuevas_molecular_2017,schinabeck_hierarchical_2018,schinabeck_hierarchical_2020} energy and charge transfer in molecular aggregates,\cite{ishizaki_unified_2009,may_charge_2011,kreisbeck_high-performance_2011,chin_role_2013,barford_electronic_2013,lee_semiclassical_2016,jang_delocalized_2018} and relaxation dynamics in condensed-phase molecular systems.\cite{tanimura_time_1989,ishizaki_quantum_2005,nitzan_chemical_2006,tanimura_numerically_2020} Achieving highly accurate simulations of open quantum systems across experimentally relevant parameter scales continues to be demanding because many realistic setups operate in strongly coupled, non-perturbative and non-Markovian regimes. In these regimes, standard approximations fail because the strong system-environment interaction prevents a perturbative expansion, and memory effects may become dominant, requiring an explicit tracking of the state of environmental degrees of freedom rather than treating the bath as a memoryless reservoir. The numerical complexity is further enhanced if the system of interest is itself a strongly correlated many-body problem.

Consequently, sophisticated numerical approaches are required to describe open quantum systems that operate in strongly coupled, non-perturbative, and non-Markovian regimes. Depending on the physical setting, commonly used methods include wave-function-based approaches such as the multi-configurational time-dependent Hartree method (MCTDH)\cite{meyer_multi-configurational_1990,manthe_wavepacket_1992} and its multi-layer extension,\cite{wang_multilayer_2003,manthe_multilayer_2008,wang_multilayer_2015,wang_numerically_2009,vendrell_multilayer_2011,wang_numerically_2011,wang_numerically_2013,wang_employing_2016,wang_multilayer_2018,larsson_tensor_2024} non-equilibrium Green's function (NEGF) techniques,\cite{mitra_phonon_2004,galperin_resonant_2006,ryndyk_nonequilibrium_2006,hartle_multimode_2008,novotny_nonequilibrium_2011,erpenbeck_effect_2015} path integral methods,\cite{feynman_theory_1963,caldeira_path_1983,makri_tensor_1995,muhlbacher_real-time_2008,weiss_iterative_2008,segal_numerically_2010,cohen_taming_2015,lee_semiclassical_2016,strathearn_efficient_2017,jorgensen_exploiting_2019,bose_pairwise_2022} and a variety of quantum master equation approaches.\cite{nakajima_quantum_1958,zwanzig_ensemble_1960,gurvitz_microscopic_1996,li_quantum_2005,breuer_theory_2007,hartle_resonant_2011} Among the available approaches to open quantum system dynamics, the hierarchical equations of motion (HEOM) approach has emerged as a powerful framework because it allows for the treatment of strong, non-Markovian system-environment interactions as well as strong correlations within the system of interest. To achieve this, the HEOM approach couples the reduced density operator of the system to a series of so-called auxiliary density operators (ADOs) in several hierarchical layers, which systematically incorporate the non-Markovian influence of the environment on the system and higher-order system-environment correlations. The method was originally proposed by Tanimura and Kubo\cite{tanimura_time_1989} to study relaxation dynamics of molecular systems coupled to an environment of harmonic oscillators. Later on it was extended to various other setups including dissipative electronic systems and electron transport.\cite{jin_dynamics_2007,jin_exact_2008,zheng_dynamic_2008,li_hierarchical_2012,hartle_decoherence_2013,zheng_kondo_2013,hartle_transport_2015,schinabeck_hierarchical_2016,schinabeck_hierarchical_2018,erpenbeck_current-induced_2020}

The main limitation of the HEOM approach remains the size of the quantum system of interest. Since typical calculations require dozens to hundreds of ADOs to be taken into account, each with the dimensionality of the reduced system density matrix, the computational cost scales poorly with the size of the system Hilbert space, posing a substantial challenge for simulations of realistic, experimentally relevant systems. To push the boundaries of numerical accessibility, several approaches to rewrite the HEOM in a more efficient representation have been proposed in the past decade. Particularly promising approaches are based on tensor-network techniques such as by employing matrix-product states (MPS) \cite{shi_efficient_2018,yan_new_2020,borrelli_expanding_2021,ke_hierarchical_2022} or tree-tensor networks (TTN) \cite{yan_efficient_2021,ke_tree_2023} to efficiently represent the set of ADOs. Furthermore, an MCTDH-like framework has been applied to HEOM in very recent works,\cite{chen_tree_2025,chen_comparison_2026} though only considering bosonic environments.

In the following, we extend these ideas to the treatment of fermionic environments consisting of one or multiple macroscopic metallic electrodes which act as heat baths and electron reservoirs. This type of environment appears for example in the study of electron transport through nanojunctions \cite{schinabeck_hierarchical_2016,schinabeck_hierarchical_2018,erpenbeck_current-induced_2020} and in the study of scattering processes of molecules off metal surfaces.\cite{preston_nonadiabatic_2025} In our numerical framework, we build upon the existing twin-space formulation of HEOM \cite{borrelli_expanding_2021,ke_hierarchical_2022} and introduce a novel approach that employs the multi-layer MCTDH method\cite{meyer_multi-configurational_1990,manthe_wavepacket_1992,wang_multilayer_2003} for the time propagation of the resulting fermionic HEOM. We benchmark our approach using the single-site Holstein model, where reference data are available.\cite{schinabeck_hierarchical_2016} Furthermore, we demonstrate the utility of our approach by presenting the first fully quantum electron transport calculations for a more demanding vibronic coupling nanojunction model with strong molecule-electrode coupling that exhibits vibrational instability,\cite{lu_blowing_2010,mack_vibrational_2025} which presented a significant challenge for previous formulations of the HEOM approach.

The remainder of this paper is organized as follows. In \cref{sec:method}, we introduce our model and summarize the fermionic HEOM formalism and its existing reformulation in the extended twin space. We describe the mapping that enables the propagation of the HEOM using the multi-layer MCTDH method, and provide some implementation details. In \cref{sec:results}, we demonstrate the proposed HEOM+MCTDH approach for two example systems. We first show benchmark calculations for a single-site Holstein-type molecular nanojunction model, and secondly, we present the first fully quantum calculations for a molecular nanojunction model with strong molecule-electrode coupling that exhibits vibrational instability. Finally, \cref{sec:conclusion} concludes and outlines future directions of research.

\section{Method}\label{sec:method}

In the following, we describe our proposed HEOM+MCTDH approach in three parts. In \cref{subsec:model}, we introduce our generic model. In \cref{subsec:heom}, we briefly sketch the derivation of the fermionic HEOM, and we describe the reformulation of the HEOM in the so-called extended twin space. In \cref{subsec:mctdh}, we describe the mapping that enables the propagation of the HEOM using the multi-layer MCTDH method, and provide some implementation details for the proposed HEOM+MCTDH approach.

\subsection{Model}\label{subsec:model}

We consider a quantum system of interest that is coupled to any number of electronic electrodes, which act as its environment. The Hamiltonian of the composite setup, therefore, consists of the system Hamiltonian, $\Hsys$, the environment, $\Hbath$, and their interaction, $\HSB$,
\begin{align}
H&=\Hsys+\Hbath+\HSB. \label{eq:total_hamiltonian}
\end{align}

The system of interest is described in second quantization using electronic creation and annihilation operators. The operator $d_i^\dagger$ ($d_i$) creates (annihilates) an electron at the $i$th electronic site in the system. The system may further comprise additional internal degrees of freedom, such as nuclear vibrational degrees of freedom. These are described in second quantization using bosonic ladder operators, $a_\alpha^\dagger$ and $a_\alpha$. For the sake of interpretability, we will occasionally employ a dimensionless position-momentum representation, 
\begin{align}
x_\alpha=\frac{1}{\sqrt{2}}(a_\alpha^\dagger+a_\alpha),\quad\quad p_\alpha=\frac{\mi}{\sqrt{2}}(a_\alpha^\dagger-a_\alpha).
\end{align} In principle, the standard fermionic HEOM and HEOM+MCTDH approach presented below can be applied to an arbitrary system Hamiltonian, $\Hsys$, that conserves the total number of electrons. In the present work, we consider two vibronic-coupling models that share the following specific structure for the system Hamiltonian,
\begin{align}
\Hsys&=\sum_{ij}\varepsilon_{ij}d_i^\dagger d_j+\sum_{\alpha}\omega_{\alpha}a_{\alpha}^\dagger a_{\alpha} \nonumber \\
&\quad+\sum_{\alpha}\sum_{ij}\lambda_{ij}^{(\alpha)}\left(a_{\alpha}+a_{\alpha}^\dagger\right)d_i^\dagger d_j, \label{eq:system_hamiltonian}
\end{align} where $\varepsilon_{ii}$ are the electronic on-site energies, $\varepsilon_{i\neq j}$ denote the hopping energies between electronic sites, $\omega_\alpha$ are the vibrational energies and $\lambda_{ij}^{(\alpha)}$ denote the linear vibronic couplings.

The electrodes are modeled as reservoirs of non-interacting electrons which act as heat baths of fixed temperature, $T_K$, and particle reservoirs of fixed chemical potential, $\mu_K$, where $K$ is an index labeling the individual electrodes. Each electrode is described in second quantization by an infinite set of electronic sites. The operator $c_{k_K}^\dagger$ ($c_{k_K}$) creates (annihilates) an electron at site $k_K$ in electrode $K$. The total environment Hamiltonian is
\begin{align}
\Hbath&=\sum_KH_K=\sum_K\sum_{k_K}\varepsilon_{k_K}c^\dagger_{k_K}c_{k_K}. \label{eq:bath_hamiltonian}
\end{align} In the context of nanojunctions investigated later in this work, the index $K$ refers to left and right electrodes, $K\in\{\mathrm{L},\mathrm{R}\}$. Non-equilibrium charge transport is induced by applying a bias voltage $\Phi$ across the junction, which in this work is applied symmetrically, such that $\mu_\mathrm{L}=-\mu_\mathrm{R}=\ec\Phi/2$, where $\ec$ is the elementary charge.

The coupling between system and environment is modeled as bilinear hopping terms, $c^\dagger_{k_K}d_i+d_i^\dagger c_{k_K}$, which allow electrons to hop from state $i$ in the system to state $k_K$ in electrode $K$ and vice versa. The coupling strengths are denoted by $V^{(i)}_{k_K}$, which in this work are assumed to be real. The coupling Hamiltonian thus takes the form
\begin{align}
\HSB&=\sum_i\sum_K\sum_{k_K}V^{(i)}_{k_K}\left(c^\dagger_{k_K}d_i+d_i^\dagger c_{k_K}\right). \label{eq:sb_hamiltonian}
\end{align} The influence of the electrodes on the system is also characterized by the spectral density functions,
\begin{align}
\Gamma^{(ij)}_K(\varepsilon)&=2\pi\sum_{k_K}V^{(i)}_{k_K}V^{(j)}_{k_K}\delta(\varepsilon_{k_K}-\varepsilon). \label{eq:spectral_density_function}
\end{align}

Because of the non-interacting nature of the electrodes and the bilinear coupling between system and environment, after tracing out the environment, the influence of the environment on the system is entirely described by two-time environment-correlation functions,
\begin{align}
C_{K,ij}^\sigma(\tau_1-\tau_2)=\sum_{k_K}V^{(i)}_{k_K}V^{(j)}_{k_K}\tr_\mathrm{B}\{c_{k_K}^\sigma(\tau_1)\,c_{k_K}^{\bar{\sigma}}(\tau_2)\,\rho_\mathrm{B}(0)\},
\end{align} where have introduced the notation $\sigma\in\{+,-\}$, $c_{k_K}^-=c_{k_K}$ and $c_{k_K}^+=c_{k_K}^\dagger$, and the interaction picture of the environment, $c_{k_K}^\sigma(t)=\me^{\mi\Hbath t/\hbar}c_{k_K}^\sigma\me^{-\mi\Hbath t/\hbar}$. We have further assumed that, at time $t=0$, the total density matrix $\rho(t)$ factorizes as
\begin{align}
\rho(0)=\rhosys(0)\otimes\rho_\mathrm{B}(0),
\end{align} with the initial state of the environment given as a product of Gibbs states,
\begin{align}
\rho_\mathrm{B}(0)=\prod_K\frac{\me^{-(H_K-\mu_KN_K)/\kB T_K}}{\tr_K\{\me^{-(H_K-\mu_KN_K)/\kB T_K}\}},
\end{align} where $N_K=\sum_{k_K}c_{k_K}^\dagger c_{k_K}$ is the number operator of electrode $K$.

The correlation functions can further be expressed via the spectral density functions as
\begin{align}
C_{K,ij}^\sigma(t)=\frac{1}{2\pi}\int_{-\infty}^\infty\diff{\varepsilon}\,\Gamma_K^{(ij)}(\varepsilon)f_K^\sigma(\varepsilon)\me^{\sigma\mi\varepsilon t/\hbar}, \label{eq:correlation_function_integral}
\end{align} where $f_K^\sigma(\varepsilon)$ is the Fermi-Dirac function,
\begin{align}
f_K^\sigma(\varepsilon)=\frac{1}{1+\me^{\sigma(\varepsilon-\mu_K)/\kB T_K}}.
\end{align} In order to derive the HEOM, the correlation functions are expanded as a sum of exponentials,
\begin{align}
C_{K,ij}^\sigma(t)=V_K^{(i)}V_K^{(j)}\sum_{\ell=0}^{\ell_\mathrm{max}}\eta_{K,\sigma,\ell}\me^{-\gamma_{K,\sigma,\ell}t}. \label{eq:correlation_function_expansion}
\end{align} To obtain this expansion in practice, we employ a Padé decomposition scheme \cite{hu_communication_2010,hu_pade_2011} on the integral in \cref{eq:correlation_function_integral}. Note that here we have introduced new coupling constants, $V^{(i)}_K\in\mathbbm{R}$, which are fixed for a given electrode $K$ and do not depend on the intra-electrode indices $k_K$.

\subsection{HEOM in the extended twin space}\label{subsec:heom}

\subsubsection{Standard fermionic HEOM}

In the following section, we briefly sketch the standard fermionic HEOM. For a more detailed derivation, we refer the reader to Refs.~\onlinecite{jin_exact_2008,hartle_decoherence_2013,ye_heom-quick_2016,schinabeck_hierarchical_2018}.

The central idea of the HEOM approach is to exploit the exponential expansion of environment-correlation functions from \cref{eq:correlation_function_expansion}. The self-similarity of the exponential terms with respect to a time-derivative allow one to define a set of auxiliary density operators (ADOs), $\rho^{\mathbf{n}}(t)$, which are utilized to systematically incorporate the non-Markovian and higher-order influence of the environment on the system in a non-perturbative and hierarchical manner. For the setup and assumptions introduced in \cref{subsec:model}, the equation of motion for an ADO is \cite{jin_exact_2008,ke_hierarchical_2022}
\begin{align}
\dpar{t}\rho^{\mathbf{n}}(t)&=-\frac{\mi}{\hbar}\left[\Hsys,\rho^{\mathbf{n}}(t)\right]-\left(\sum_r n_r\gamma_r\right)\rho^{\mathbf{n}}(t) \nonumber \\
&\quad-\frac{\mi}{\hbar}\sum_r\delta_{n_r,1}(-1)^{\sum\limits_{s<r}n_s}\mathcal{C}_r\rho^{\mathbf{n}-\mathbf{1}_r}(t) \label{eq:heom} \\
&\quad-\frac{\mi}{\hbar}\sum_r\delta_{n_r,0}(-1)^{\sum\limits_{s<r}n_s}\mathcal{A}_{\bar{r}}\rho^{\mathbf{n}+\mathbf{1}_r}(t). \nonumber
\end{align} These are the so-called hierarchical equations of motion (HEOM). In the above, we have introduced the vector-valued index
\begin{align}
\mathbf{n}=(n_1,n_2,\dots)
\end{align} to label the ADOs. Since this is a fermionic hierarchy, each index can be $n_r\in\{0,1\}$. The index $r$ is a combined index that runs over all possible combinations of $K$, $\sigma$ and $\ell$ as defined in \cref{eq:correlation_function_expansion}, i.e., $\gamma_r=\gamma_{K_r,\sigma_r,\ell_r}$ and $\eta_r=\eta_{K_r,\sigma_r,\ell_r}$. For $\mathbf{n}=\mathbf{0}=(0,0,\dots)$, we recover the reduced density operator of the system, $\rho^{\mathbf{n}=\mathbf{0}}(t)=\rhosys(t)$. The superoperators, $\mathcal{A}_{\bar{r}}$ and $\mathcal{C}_r$, are defined as
\begin{align}
\mathcal{A}_{\bar{r}}\rho^{\mathbf{n}}(t)&=\sum_iV^{(i)}_{K_r}\Big[d_i^{\bar{\sigma}_r}\rho^{\mathbf{n}}(t)+(-)^{\sum\limits_sn_s}\rho^{\mathbf{n}}(t)d_i^{\bar{\sigma}_r}\Big],\\
\mathcal{C}_r\rho^{\mathbf{n}}(t)&=\sum_iV^{(i)}_{K_r}\Big[\eta_rd_i^{\sigma_r}\rho^{\mathbf{n}}(t)-(-)^{\sum\limits_sn_s}\eta_{\bar{r}}^\ast\rho^{\mathbf{n}}(t)d_i^{\sigma_r}\Big].
\end{align} They couple each ADO to the next higher (lower) tier where one more (fewer) of the $n_r$ is occupied,
\begin{align}
\mathbf{n}\pm\mathbf{1}_r=(n_1,n_2,\dots,n_r\pm 1,\dots).
\end{align}

The HEOM in \cref{eq:heom} form a closed set of coupled first-order ordinary differential equations for the ADOs, $\rho^{\mathbf{n}}(t)$, which can in principle be time-integrated or solved directly for the steady state using standard numerical techniques. In practical applications, the index $r$ usually spans $10-100$ combinations of $(K_r,\sigma_r,\ell_r)$. The full hierarchy thus typically contains far more than $2^{10}$ ADOs, each of which is a linear operator with the same dimensions as the reduced system density operator. Although the hierarchy can often be truncated at the second or third tier when the system-environment coupling is not too strong, the truncated hierarchy typically still contains between several dozen and several hundred ADOs.

The application of standard HEOM is therefore often limited to systems of moderate complexity. This limitation is particularly severe for systems with many vibrational degrees of freedom, since their representation rapidly increases the dimensionality of each ADO. Tensor-network techniques provide a natural way to address this dimensionality problem by compressing the information stored in the ADOs, for example using matrix-product states (MPS) \cite{shi_efficient_2018,yan_new_2020,borrelli_expanding_2021,ke_hierarchical_2022} or tree-tensor networks (TTN).\cite{yan_efficient_2021,ke_tree_2023} The HEOM+MCTDH approach proposed in this paper follows a related strategy: the set of ADOs is represented in a tree structure and propagated directly using multi-layer MCTDH. To this end, we first reformulate the HEOM in \cref{eq:heom} in the more compact extended twin-space representation.\cite{ke_hierarchical_2022}

\subsubsection{Transformation to the extended twin space}

The ADOs, $\rho^{\mathbf{n}}(t)$, are linear operators in the vector space of endomorphisms on the Hilbert space of the system, $\operatorname{End}(\mathcal{H}_\mathrm{S})$. This space of endomorphisms is isomorphic to the tensor space of $\mathcal{H}_\mathrm{S}$ with its dual space $\mathcal{H}_\mathrm{S}^\dagger$, $\operatorname{End}(\mathcal{H}_\mathrm{S})\cong\mathcal{H}_\mathrm{S}\otimes\mathcal{H}_\mathrm{S}^\dagger$. This is generally exploited to expand an operator $A\in\operatorname{End}(\mathcal{H}_\mathrm{S})$ as
\begin{align}
A=\sum_{m,n}A_{mn}\ket{m}\!\!\bra{n},
\end{align} for some arbitrary orthonormal basis $\{\ket{m}\}_m$ of $\mathcal{H}_\mathrm{S}$, where the $\otimes$ notation has been omitted as usual. Since $\mathcal{H}_\mathrm{S}$ is canonically isomorphic to its dual space $\mathcal{H}^\dagger_\mathrm{S}$, we can equivalently write the operator $A$ as a vector in the tensor space $\mathcal{H}_\mathrm{S}\otimes\mathcal{H}_\mathrm{S}$. The second factor is usually written with a tilde for distinguishability, $\operatorname{End}(\mathcal{H}_\mathrm{S})\cong\mathcal{H}_\mathrm{S}\otimes\tilde{\mathcal{H}}_\mathrm{S}$. The tensor space $\mathcal{H}_\mathrm{S}\otimes\tilde{\mathcal{H}}_\mathrm{S}$ is well-known as the twin space from thermo-field dynamics.\cite{suzuki_thermo_1985,suzuki_density_1991,borrelli_density_2019,borrelli_finite_2021,borrelli_expanding_2021,ke_hierarchical_2022}

In the twin space, we define the embedded action of $A\in\operatorname{End}(\mathcal{H}_\mathrm{S})$ on any vector $\dblket{B}\in\mathcal{H}_\mathrm{S}\otimes\tilde{\mathcal{H}}_\mathrm{S}$ as
\begin{align}
A\dblket{B}:=(A\otimes\tilde{\mathbbm{1}})\dblket{B},
\end{align} and, analogously, the embedded action of $\tilde{A}\in\operatorname{End}(\tilde{\mathcal{H}}_\mathrm{S})$ on any $\dblket{B}\in\mathcal{H}_\mathrm{S}\otimes\tilde{\mathcal{H}}_\mathrm{S}$ as
\begin{align}
\tilde{A}\dblket{B}:=\big(\mathbbm{1}\otimes\tilde{A}\big)\dblket{B}.
\end{align} A central component of twin-space theory is the so-called identity vector,
\begin{align}
\dblket{\mathbbm{1}}:=\sum_m\ket{m}\otimes\ket{\tilde{m}}, \label{eq:twin_space_identity_vector}
\end{align} where $\{\ket{m}\}_m$ and $\{\ket{\tilde{m}}\}_{\tilde{m}}$ are mutually corresponding orthonormal bases of $\mathcal{H}_\mathrm{S}$ and $\tilde{\mathcal{H}}_\mathrm{S}$, respectively. The identity vector serves to map any $A\in\operatorname{End}(\mathcal{H}_\mathrm{S})$ to its corresponding ket vector $\dblket{A}\in\mathcal{H}_\mathrm{S}\otimes\tilde{\mathcal{H}}_\mathrm{S}$,
\begin{align}
A\dblket{\mathbbm{1}}
&=\sum_{m,n}A_{mn}\ket{m}\otimes\ket{\tilde{n}}=\dblket{A}. \label{eq:twin_space_translation}
\end{align} The identity vector further maps $\dblket{A}\in\mathcal{H}_\mathrm{S}\otimes\tilde{\mathcal{H}}_\mathrm{S}$ to a corresponding operator $\tilde{A}^\dagger\in\operatorname{End}(\tilde{\mathcal{H}}_\mathrm{S})$,
\begin{align}
\dblket{A}&=\sum_{m,n}A_{mn}\ket{m}\otimes\ket{\tilde{n}} \nonumber \\
&=\sum_{m,n}(A^\dagger)^\ast_{nm}\ket{m}\otimes\ket{\tilde{n}}=\tilde{A}^\dagger\dblket{\mathbbm{1}}. \label{eq:twin_space_translation_2}
\end{align} Note that the corresponding operator in the tilde space contains a complex conjugation,
\begin{align}
\tilde{A}^\dagger&:=\sum_{m,n}(A^\dagger)^\ast_{mn}\ket{\tilde{m}}\!\!\bra{\tilde{n}}.
\end{align}

The advantage of the twin-space formulation becomes apparent when considering products of operators. Given a linear operator, $A\in\operatorname{End}(\mathcal{H}_\mathrm{S})$, acting on an ADO, $\rho\in\operatorname{End}(\mathcal{H}_\mathrm{S})$. Then, left multiplication is represented straightforwardly as
\begin{align}
A\rho\dblket{\mathbbm{1}}=A\dblket{\rho}.
\end{align} But importantly, right multiplication is mapped to
\begin{align}
\rho A\dblket{\mathbbm{1}}\overset{(\ref{eq:twin_space_translation})}{=}\rho\dblket{A}\overset{(\ref{eq:twin_space_translation_2})}{=}\rho\tilde{A}^\dagger\dblket{\mathbbm{1}}=\tilde{A}^\dagger\rho\dblket{\mathbbm{1}}=\tilde{A}^\dagger\dblket{\rho},
\end{align} where $\rho$ and $\tilde{A}^\dagger$ commute because they act in the different factor spaces, $\mathcal{H}_\mathrm{S}$ and $\tilde{\mathcal{H}}_\mathrm{S}$. This implies, in particular, that we can write commutators as differences,
\begin{align}
\left[A,\rho\right]\dblket{\mathbbm{1}}=\left(A-\tilde{A}^\dagger\right)\dblket{\rho}.
\end{align}

Exploiting the above, the HEOM can be rewritten in twin space as a hierarchy of coupled vector-valued differential equations. To consolidate this hierarchy into a single equation, we map the vector-valued index $\mathbf{n}=(n_1,n_2,\dots)$ to a Fock space of virtual fermionic modes,\cite{borrelli_density_2019,ke_hierarchical_2022} $\ket{\mathbf{n}}=\ket{n_1,n_2,\dots}$, and introduce for each index $r$ two pairs of creation and annihilation operators, which create or annihilate the respective fermion from the left and right,
\begin{align}
\mathfrak{c}_r^{<,\dagger}\ket{\mathbf{n}}&=\delta_{n_r,0}(-1)^{\sum\limits_{s<r}n_s}\ket{\mathbf{n}+\mathbf{1}_r}, \\
\mathfrak{c}_r^<\ket{\mathbf{n}}&=\delta_{n_r,1}(-1)^{\sum\limits_{s<r}n_s}\ket{\mathbf{n}-\mathbf{1}_r}, \\
\mathfrak{c}_r^{>,\dagger}\ket{\mathbf{n}}&=\delta_{n_r,0}(-1)^{\sum\limits_{s>r}n_s}\ket{\mathbf{n}+\mathbf{1}_r}, \\
\mathfrak{c}_r^>\ket{\mathbf{n}}&=\delta_{n_r,1}(-1)^{\sum\limits_{s>r}n_s}\ket{\mathbf{n}-\mathbf{1}_r}.
\end{align} We can now form the tensor product of the vectorized ADOs with their respective virtual mode state,
\begin{align}
\dblket{\rho^{\mathbf{n}}(t)}\otimes\ket{\mathbf{n}},
\end{align} and sum all terms to a consolidated state vector,
\begin{align}
\ket{\Psi(t)}=\sum_{n_1,n_2,\dots}\dblket{\rho^{\mathbf{n}}(t)}\otimes\ket{\mathbf{n}}. \label{eq:heom_condensed_state_vector}
\end{align}
Taking the time-derivative, the HEOM can be reformulated as a single Schrödinger-type equation,
\begin{align}
\mi\hbar\dpar{t}\ket{\Psi(t)}=\mathscr{H}\ket{\Psi(t)}, \label{eq:heom_condensed}
\end{align}
where the non-Hermitian super-Hamiltonian $\mathscr{H}$ is
\begin{align}
\mathscr{H}&=\Hsys-\tilde{\Hsys}-\mi\hbar\sum_r\gamma_r\mathfrak{c}_r^\dagger\mathfrak{c}_r \nonumber \\
&\quad+\sum_{r,i}V^{(i)}_{K_r}\left(\eta_rd_i^{\sigma_r}\mathfrak{c}_r^{<,\dagger}-\eta_{\bar{r}}^\ast\tilde{d}_i^{ \bar{\sigma}_r}\mathfrak{c}_r^{>,\dagger}\right) \label{eq:super_hamiltonian} \\
&\quad+\sum_{r,i}V^{(i)}_{K_r}\left(d_i^{\bar{\sigma}_r}\mathfrak{c}_r^<-\tilde{d}_i^{\sigma_r}\mathfrak{c}_r^>\right). \nonumber
\end{align}

\subsubsection{Wide-band limit}

In this paper, all results are calculated in the wide-band limit,
\begin{align}
\Gamma^{(ij)}_K(\varepsilon)=\mathrm{const.}=2\pi V^{(i)}_KV^{(j)}_K.
\end{align} In this particular case, the integral in \cref{eq:correlation_function_integral} diverges at $t=0$, and we obtain a modified expansion for the correlation functions including a $\delta$-function,
\begin{align}
C_{K,ij}^\sigma(t)=V_K^{(i)}V_K^{(j)}\Big(\pi\hbar\delta(t)+\sum_{\ell=0}^{\ell_\mathrm{max}}\eta_{K,\sigma,\ell}\me^{-\gamma_{K,\sigma,\ell}t}\Big). \label{eq:correlation_function_expansion_WBL}
\end{align} Due to the $\delta$-function, an additional term appears in the HEOM. In the extended twin-space formulation, we still obtain a Schrödinger-type equation,
\begin{align}
\mi\hbar\dpar{t}\ket{\Psi(t)}=\mathscr{H}_\mathrm{WBL}\ket{\Psi(t)}, \label{eq:heom_condensed_wbl}
\end{align}
though we also find corresponding additional terms in the super-Hamiltonian,\cite{ke_hierarchical_2022}
\begin{align}
\mathscr{H}_\mathrm{WBL}&=\mathscr{H}-\mi\pi\sum_{K,i}\big(V_K^{(i)}\big)^2 \nonumber \\
&\quad+\mi\pi\sum_{K,\sigma,i,j}V_K^{(i)}V_K^{(j)}d_i^{\bar{\sigma}}\tilde{d}_j^{\bar{\sigma}}\mathbf{\Sigma}, \label{eq:super_hamiltonian_wbl}
\end{align} where
\begin{align}
\mathbf{\Sigma}\ket{\mathbf{n}}&=(-1)^{\sum\limits_{s}n_s}\ket{\mathbf{n}}.
\end{align}

\subsubsection{Observables of interest}\label{subsec:observables}

Given that the HEOM approach generates an equation of motion for the reduced density operator of the system, $\rhosys(t)=\rho^{\mathbf{n}=\mathbf{0}}(t)$, it provides direct access to the expectation value of any system observable $O$ via $\tr\{O\rhosys(t)\}$. Additional observables that are functions of the bath operators, such as the electric current or heat current, are available through the ADOs. In this work, the main observables of interest are the expectation values of the electric current, $\langle I_K\rangle$, the electronic populations, $\langle d_i^\dagger d_i\rangle$, and the vibrational excitations, $\langle a_\alpha^\dagger a_\alpha\rangle$.

At any time $t$, we can extract the reduced density operator from the consolidated state vector as
\begin{align}
\dblket{\rhosys(t)}=\dblket{\rho^{\mathbf{n}=\mathbf{0}}(t)}=\braket{\mathbf{0}|\Psi(t)}.
\end{align} In twin space, the trace of an operator takes the form of a scalar product,
\begin{align}
\tr\{A\}=\dblbra{\mathbbm{1}}A\dblket{\mathbbm{1}}=\dblbraket{\mathbbm{1}|A}.
\end{align} Therefore, we find the electronic populations and vibrational excitations via
\begin{align}
\langle d_i^\dagger d_i\rangle&=\dblbra{\mathbbm{1}}d_i^\dagger d_i\dblket{\rhosys(t)}=\dblbra{\mathbbm{1}}d_i^\dagger d_i\braket{\mathbf{0}|\Psi(t)}, \label{eq:occupation} \\
\langle a_i^\dagger a_i\rangle&=\dblbra{\mathbbm{1}}a_i^\dagger a_i\dblket{\rhosys(t)}=\dblbra{\mathbbm{1}}a_i^\dagger a_i\braket{\mathbf{0}|\Psi(t)}. \label{eq:excitation}
\end{align}

The electric current is an observable of the first-tier ADOs \cite{jin_exact_2008} and can be obtained as
\begin{align}
\left\langle I_K\right\rangle&=\frac{\mi\ec}{\hbar}\sum_{\overset{r}{K_r=K}}\sum_iV^{(i)}_{K}\dblbra{\mathbbm{1}}\sigma_r d_i^{\bar{\sigma}_r}\braket{\mathbf{1}_r|\Psi(t)}. \label{eq:electric_current}
\end{align} In the wide-band limit, the expression for the electric current takes the longer form
\begin{align}
\left\langle I_K\right\rangle&=\frac{\mi\ec}{\hbar}\sum_{\overset{r}{K_r=K}}\sum_iV^{(i)}_{K}\dblbra{\mathbbm{1}}\sigma_r d_i^{\bar{\sigma}_r}\braket{\mathbf{1}_r|\Psi(t)} \nonumber \\
&\quad+\frac{2\pi\ec}{\hbar}\sum_{i,j,\sigma}V_K^{(i)}V_K^{(j)}\dblbra{\mathbbm{1}}\sigma d_i^{\bar{\sigma}} d_j^\sigma\braket{\mathbf{0}|\Psi(t)}, \label{eq:electric_current_wbl}
\end{align}
due to the additional terms in the super-Hamiltonian.

\subsection{Time propagation using MCTDH}\label{subsec:mctdh}

\subsubsection{MCTDH approach to HEOM}

The key novelty of this paper is to employ the multi-layer MCTDH approach \cite{meyer_multi-configurational_1990,manthe_wavepacket_1992,wang_multilayer_2003,manthe_multilayer_2008} to propagate the HEOM from \cref{eq:heom_condensed,eq:heom_condensed_wbl}. The multi-layer MCTDH method provides an efficient framework for accurate multi-dimensional quantum dynamics simulations. To apply the MCTDH approach to the HEOM, first the set of ADOs, $\{\rho^{\mathbf{n}}(t)\}$, or equivalently, the consolidated state vector from \cref{eq:heom_condensed_state_vector} must be mapped onto a Hilbert space wavefunction, $\Psi(\mathbf{q},t)$, that depends on $f$ coordinates, $\mathbf{q}=(q_1,q_2,\dots,q_f)$. Denoting the combined set of coordinates of the system of interest with $\mathbf{x}$, the wavefunction is defined by
\begin{align}
\Psi(\mathbf{q},t)=\Psi(\mathbf{x},\mathbf{\tilde{x}},\mathbf{n},t)=\rho^\mathbf{n}(\mathbf{x},\mathbf{\tilde{x}},t)=\braket{\mathbf{\tilde{x}}|\rho^\mathbf{n}(t)|\mathbf{x}},
\end{align} where the set of coordinates $\mathbf{q}$ combines the two sets of coordinates, $\mathbf{x}$ and $\mathbf{\tilde{x}}$, present in the ADOs, and the vector-valued index $\mathbf{n}$. Note that $\mathbf{x}$ and $\mathbf{\tilde{x}}$ are independent sets of coordinates. The scalar product of two such wavefunctions, $\Psi_{(1)}$ and $\Psi_{(2)}$, is consequently given by
\begin{align}
\braket{\Psi_{(1)}|\Psi_{(2)}}=\sum_{\mathbf{n}}\tr\big\{\rho_{(1)}^{\mathbf{n}\dagger}\,\rho_{(2)}^\mathbf{n}\big\},
\end{align} and the Dirac-Frenkel-type variational principle used in the derivation of the (multi-layer) MCTDH equations of motion is
\begin{align}
0&=\bra{\delta\Psi(t)}\mi\hbar\dpar{t}-\mathscr{H}\ket{\Psi(t)} \nonumber \\
&=\sum_{\mathbf{n}}\tr\Big\{\delta\rho^{\mathbf{n}\dagger}\Big(\dpar{t}\rho^{\mathbf{n}}(t)-\mathcal{L}\rho^{\mathbf{n}}(t)\Big)\Big\}, \label{eq:dirac_frenkel}
\end{align} where the Liouvillian $\mathcal{L}\rho^{\mathbf{n}}(t)$ denotes the right hand side of \cref{eq:heom}. The present HEOM+MCTDH approach thus combines elements of the MCTDH approach to density matrix propagation,\cite{gerdts_wave_1997,raab_multiconfiguration_1999} the state-averaged MCTDH scheme,\cite{manthe_state_2008} and the MCTDH in second quantization representation (MCTDH-SQR) \cite{wang_numerically_2009} for the description of the HEOM in the MCTDH framework.

The multi-layer MCTDH approach employs a tree-shaped tensor contraction scheme to represent the wavefunction $\Psi(\mathbf{q},t)$. The tree-shaped structure is constructed by grouping the coordinates $\mathbf{q}$ into sets of logical coordinates $\mathbf{q}^{l;\kappa_1,\dots,\kappa_{l-1}}$, where the superscript $(l;\kappa_1,\dots,\kappa_{l-1})$ specifies a specific node in the tree. The top node is labeled $(1)$, the first layer nodes $(2;\kappa_1)$, where $\kappa_1$ runs over all nodes of the first layer, the second layer nodes are labeled $(3;\kappa_1,\kappa_2)$, where $\kappa_2$ runs over all second-layer nodes connected to the first-layer node $(2;\kappa_1)$, and so on. The wavefunction $\Psi(q_1,\dots,q_f,t)$ is expanded in an orthonormal set of single-particle functions (SPFs),
\begin{align}
\Psi(q_1,\dots,q_f,t)=\sum_{j_1,\dots,j_\nu}A^1_{j_1,\dots,j_\nu}(t)\prod_\kappa\phi^{1;\kappa}_{j_\kappa}(\mathbf{q}^{2;\kappa},t),
\end{align} where the single-particle functions are recursively defined by
\begin{align}
\phi^{l;\kappa_1,\dots,\kappa_l}_n=\sum_{j_1,\dots,j_\nu}A^{l+1;\kappa_1,\dots,\kappa_l}_{n;j_1,\dots,j_\nu}\prod_{\kappa_{l+1}=1}^\nu\phi^{l+1;\kappa_1,\dots,\kappa_{l+1}}_{j_{\kappa_{l+1}}}.
\end{align} Here, the $A^{l+1;\kappa_1,\dots,\kappa_l}(t)$ denote time-dependent expansion coefficients, and $\nu$ is the number of edges connecting the node $(l;\kappa_1,\dots,\kappa_{l-1})$ to lower nodes $(l+1;\kappa_1,\dots,\kappa_l)$. At the bottom layer of the tree, the SPFs are replaced by time-independent basis functions $\chi^\kappa_n$.

Employing the Dirac-Frenkel variational principle from \cref{eq:dirac_frenkel} and the gauge condition
\begin{align}
0=\braket{\phi^{l;\kappa_1,\dots,\kappa_l}_n|\dpar{t}\phi^{l;\kappa_1,\dots,\kappa_l}_m},
\end{align} one can derive the (multi-layer) MCTDH equations of motion
\begin{align}
\mi\hbar\dpar{t}A^1_{i_1,\dots,i_\nu}&=\sum_{j_1,\dots,j_\nu}\braket{\prod_{\kappa=1}^\nu\phi^{1;\kappa}_{i_\kappa}|\mathscr{H}|\prod_{\kappa=1}^\nu\phi^{1;\kappa}_{j_\kappa}}A^1_{j_1,\dots,j_\nu}, \\
\mi\hbar\dpar{t}\phi^{l;\kappa_1,\dots,\kappa_l}&=\big(\mathbbm{1}-P^{l;\kappa_1,\dots,\kappa_l}_{\kappa_l}\big) \nonumber \\
&\quad\cdot\sum_{j,m}\big(\varrho^{l;\kappa_1,\dots,\kappa_l}\big)^{-1}_{nj}\langle\mathscr{H}\rangle^{l;\kappa_1,\dots,\kappa_l}_{jm}\phi^{l;\kappa_1,\dots,\kappa_l}_m.
\end{align} The single-particle density matrices and mean-field matrices in the above equations are given by
\begin{align}
\varrho^{l;\kappa_1,\dots,\kappa_l}_{nm}=\braket{\Psi^{l;\kappa_1,\dots,\kappa_l}_n|\Psi^{l;\kappa_1,\dots,\kappa_l}_m}
\end{align} and
\begin{align}
\langle\mathscr{H}\rangle^{l;\kappa_1,\dots,\kappa_l}_{nm}=\braket{\Psi^{l;\kappa_1,\dots,\kappa_l}_n|\mathscr{H}|\Psi^{l;\kappa_1,\dots,\kappa_l}_m},
\end{align} where the single-hole functions $\Psi^{l;\kappa_1,\kappa_2,\dots,\kappa_l}_i$ are defined via
\begin{align}
\Psi=\sum_n\Psi^{l;\kappa_1,\kappa_2,\dots,\kappa_l}_n\cdot\phi^{l;\kappa_1,\kappa_2,\dots,\kappa_l}_n.
\end{align} The projector on the space spanned by the SPFs of the respective layer is defined by
\begin{align}
P^{l;\kappa_1,\dots,\kappa_l}_{\kappa_l}=\sum_m\ket{\phi^{l;\kappa_1,\dots,\kappa_l}_m}\!\!\bra{\phi^{l;\kappa_1,\dots,\kappa_l}_m}.
\end{align}

An efficient implementation of the multi-layer MCTDH approach requires an efficient scheme for the calculations of the mean-field matrices at each node of the tree. As shown in Ref.~\onlinecite{manthe_multilayer_2008}, an efficient recursive procedure for the computation of the mean-field matrices can be implemented for Hamiltonians showing a sum-of-products form,
\begin{align}
\mathscr{H}=\sum_n\Big(c_n\prod_{k=1}^f\hat{h}^{\mathrm{prim};\kappa}_n\Big),
\end{align} where $\hat{h}^{\mathrm{prim};\kappa}_n$ denotes a single-particle operator which acts only on the coordinate $q_\kappa$. Inspection of \cref{eq:super_hamiltonian,eq:super_hamiltonian_wbl} immediately reveals that the super-Hamiltonian of the HEOM indeed shows this sum-of-products form.

\subsubsection{Implementation and numerical details}\label{subsec:heom+mctdh}

For the integration of the (multi-layer) MCTDH equations of motion, a constant mean-field (CMF) scheme \cite{beck_efficient_1997,manthe_integration_2006} is employed. Consequently, there are two accuracy parameters to be specified: the standard integration accuracy, $\epsilon$, and an additional parameter, $\epsilon_\mathrm{CMF}$, which controls the re-evaluation of the mean-field matrices.

During the propagation, the natural populations, i.e., the eigenvalues of the single-particle density matrices,
\begin{align}
\varrho^{l;\kappa_1,\dots}(t)=\sum_ip^{l;\kappa_1,\dots}_i(t)\ket{p^{l;\kappa_1,\dots}_i(t)}\!\!\bra{p^{l;\kappa_1,\dots}_i(t)},
\end{align} are used to monitor the convergence with respect to the SPF basis set size. We require that the smallest natural population stays below a pre-defined threshold value $\delta$,
\begin{align}
\min_i\big\{p^{l;\kappa_1,\dots}_i(t)\big\}\overset{!}{<}\delta. \label{eq:mctdh_spf_threshold}
\end{align} As the time integration proceeds, fulfillment of this requirement is checked at pre-determined intervals, $t_\mathrm{check}$. If the threshold is exceeded for one or more SPF basis sets, the affected basis sets are extended by a fixed number of (unoccupied) basis functions. The new basis functions are chosen randomly, ensuring orthogonality to the existing basis functions. Optionally, when such a basis extension occurs, a fixed number of checking intervals, $N_\mathrm{redo}$, may be recalculated retrospectively with the extended basis to compensate for lost accuracy.

The vector-valued index $\mathbf{n}$ consists of a large number ($m$) of indices, $n_i\in\{0,1\}$. To increase the numerical efficiency, we combine the entire $\mathbf{n}$-vector into a single coordinate, $q_1=\bar{n}$, in the MCTDH calculation. The translation scheme is as follows:
\begin{itemize}
\item First, we associate $\bar{n}=0$ with $\mathbf{n}=\mathbf{0}$ and $\bar{n}\in\{1,2,\dots,m\}$ with the single excitations $\mathbf{1}_1,\mathbf{1}_2,\dots,\mathbf{1}_m$.
\item Second, we associate $\bar{n}\in\{m+1,m+2,\dots,m+\binom{m}{2}\}$ with the double excitations $\mathbf{1}_1+\mathbf{1}_2,\mathbf{1}_1+\mathbf{1}_3,\dots,\mathbf{1}_{m-1}+\mathbf{1}_m$.
\item Third, we associate $\bar{n}\in\{m+\binom{m}{2}+1,m+\binom{m}{2}+2,\dots,m+\binom{m}{2}+\binom{m}{3}\}$ with the triple excitations $\mathbf{1}_1+\mathbf{1}_2+\mathbf{1}_3,\mathbf{1}_1+\mathbf{1}_2+\mathbf{1}_4,\dots,\mathbf{1}_{m-2}+\mathbf{1}_{m-1}+\mathbf{1}_m$.
\item And so on for higher numbers of excitations.
\end{itemize} To achieve a truncation to $N$th tier, the basis of index $\bar{n}$ is truncated at a size of
\begin{align}
\sum_{n=0}^{N}\binom{m}{n}=\binom{m}{0}+\binom{m}{1}+\dots+\binom{m}{N},\ N\leq m.
\end{align} Convergence is achieved if the total population of all states in the highest included tier is negligible. All calculations in this work were truncated at either second or third tier. For certain models, it is advantageous to combine the indices in the $\mathbf{n}$-vector not into a single index but into separate indices per electrode. In the usual nanojunction case, we hereby obtain a left index, $q_1=\bar{n}_\mathrm{L}$, and a right index, $q_2=\bar{n}_\mathrm{R}$, in the MCTDH calculation. The translation scheme is analogous to above, with the exception that $\bar{n}_\mathrm{L}$ ($\bar{n}_\mathrm{R}$) only considers the virtual fermionic modes of the left (right) electrode.

While the standard HEOM approach conserves the trace of the reduced density matrix, $\tr\{\rhosys(t)\}=\dblbra{\mathbbm{1}}\braket{\mathbf{0}|\Psi(t)}$, the HEOM+MCTDH approach does so only in the limit of converged SPF basis set sizes. The HEOM+MCTDH approach shares this feature with MCTDH density matrix propagation (see Ref.~\onlinecite{gerdts_wave_1997} for a detailed discussion). Consequently, deviation of $\tr\{\rhosys(t)\}$ from $1$ can be used to measure the level of convergence achieved. This problem also affects the observables of interest. For observables of the form $\tr\{A\rhosys(t)\}$, for some operator $A\in\operatorname{End}(\mathcal{H}_\mathrm{S})$, a deviation from their true values by precisely the factor $\tr\{\rhosys(t)\}$ is expected. Since the HEOM are linear in the ADOs, for a given set of ADOs $\{\rho^\mathbf{n}(t)\}_\mathbf{n}$ that fulfill the HEOM, the set of multiples $\{\lambda\cdot\rho^\mathbf{n}(t)\}_\mathbf{n}$ will also fulfill the same HEOM. Consequently, also for observables of the general form $\tr\{A\rho^\mathbf{n}(t)\}$, the deviation from the true value is expected to be the same factor $\tr\{\rhosys(t)\}$. Therefore, we renormalize all observables with respect to $\tr\{\rhosys(t)\}$. An analogous renormalization is used, for example, in the evaluation of expectation values in the presence of complex absorbing operators.\cite{westermann_decoherence_2012}

Designing the tree structure used in the MCTDH calculation, the ability to efficiently calculate our observables of interest has to be considered. As discussed in \cref{subsec:observables}, the calculation of our observables of interest involve many scalar products of the form,
\begin{align}
\dblbra{\mathbbm{1}}A\braket{\mathbf{0}|\Psi(t)},\quad\text{and}\quad\dblbra{\mathbbm{1}}A\braket{\mathbf{1}_r|\Psi(t)},
\end{align} for some linear operator $A\in\operatorname{End}(\mathcal{H}_\mathrm{S})$. Note that the identity vector, $\dblket{\mathbbm{1}}$, defined in \cref{eq:twin_space_identity_vector}, is a highly correlated state vector, as it connects all physical degrees of freedom, $x_i$, with their corresponding ancilla, $\tilde{x}_i$. To facilitate the efficient observable calculations, it is thus desirable to place any pair of physical and ancilla coordinates, $x_i$ and $\tilde{x}_i$, at the same bottom layer node of the tree.

\section{Results}\label{sec:results}

In the following, we demonstrate the HEOM+MCTDH method for two physically relevant example systems. To verify the accuracy of the HEOM+MCTDH approach, we begin in \cref{subsec:holstein} with a benchmark for a simpler model system, where reference data are available by means of several other HEOM-based methods. As the model system, we choose a Holstein-type nanojunction model which was previously studied in Ref.~\onlinecite{schinabeck_hierarchical_2016}. Secondly, in \cref{subsec:2L2M}, we demonstrate the capabilities of the HEOM+MCTDH approach by simulating transport through a more demanding molecular nanojunction model with strong molecule-electrode coupling comprising two electronic sites and two vibrational modes. Specifically, we consider a nanojunction model that exhibits vibrational instability similar to the one studied in Ref.~\onlinecite{mack_vibrational_2025}, for which fully quantum results have not been available so far.

\subsection{Holstein model}\label{subsec:holstein}

In this section, we benchmark the HEOM+MCTDH method for a simple molecular nanojunction model that was previously studied in Refs.~\onlinecite{holstein_studies_1959,koch_franck-condon_2005,hartle_resonant_2011,schinabeck_hierarchical_2016}. The model considered is a Holstein-type model comprising one electronic site and one vibrational mode. The electronic site represents the relevant molecular orbital mediating charge transport through the nanojunction, while the vibrational mode provides an effective description of the collective influence of molecular vibrations. The electronic energy level depends linearly on the vibrational displacement, represented by $x=(a^\dagger+a)/\sqrt{2}$. The system Hamiltonian is given by
\begin{align}
\Hsys=\left(\varepsilon_0+\lambda(a+a^\dagger)\right) d^\dagger d+\omega a^\dagger a.
\end{align} As in Ref.~\onlinecite{schinabeck_hierarchical_2016}, we consider multiple parameter sets for the Holstein model---specifically, we use model 1 and model 2 as referred to in Ref.~\onlinecite{schinabeck_hierarchical_2016}. The parameters are given in \cref{tab:holstein_parameters}.

\begin{table}[H]
\centering
\begin{tabular}{|c|r|r|}
\hline
 & \multicolumn{2}{|c|}{Value [meV]} \\
Parameter & Model 1 & Model 2 \\
\hline
$\varepsilon_0$ & $300\hphantom{.0}$ & $1050\hphantom{.0}$ \\
$\varepsilon_1$ & $228\hphantom{.0}$ & $250\hphantom{.0}$ \\
$\omega$ & $200\hphantom{.0}$ & $200\hphantom{.0}$ \\
$\lambda$ & $120\hphantom{.0}$ & $400\hphantom{.0}$ \\
$\kB T$ & $25.8$ & $25.8$ \\
$\Gamma_K$ & \multicolumn{2}{|c|}{$\in\{10,100\}$} \\
\hline
\end{tabular}
\caption{\justifying Parameters used for the Holstein model 1 and model 2 as defined in Ref.~\onlinecite{schinabeck_hierarchical_2016}. $\varepsilon_1$ denotes the Lang-Firsov-transformed energy level,\cite{lang_kinetic_1963,cuevas_molecular_2017} $\varepsilon_1=\varepsilon_0-\lambda^2/\omega$.}
\label{tab:holstein_parameters}
\end{table}

In the extended twin space, we obtain the usual equation of motion for the consolidated state vector, $\ket{\Psi(t)}$, as given in \cref{eq:heom_condensed_wbl}. The (non-Hermitian) super-Hamiltonian $\mathscr{H}_\mathrm{WBL}$ is as given in \cref{eq:super_hamiltonian,eq:super_hamiltonian_wbl}. The sums over indices $i$ and $j$ disappear here since the Holstein model features a single electronic level only.

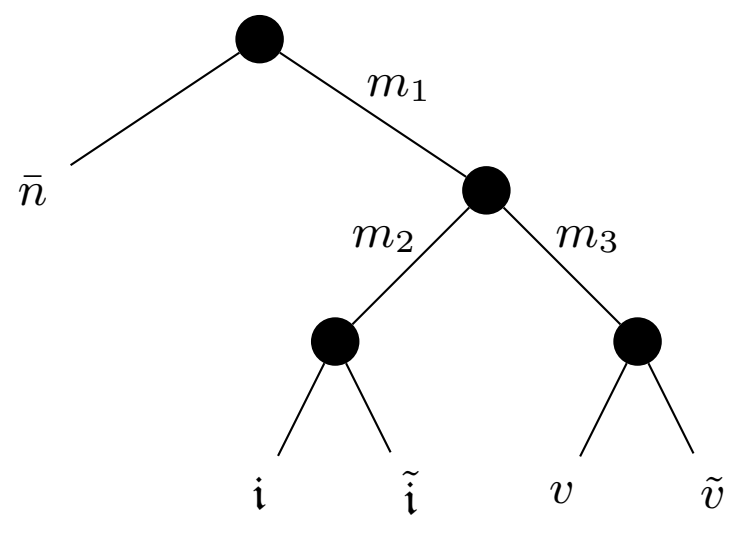
\begin{figure}[H]
\centering
\begin{tikzpicture}
        [
            grow=down,
            level 1/.style={sibling distance=3cm, level distance=1cm},
            level 2/.style={sibling distance=2cm, level distance=1cm},
            level 3/.style={sibling distance=1cm, level distance=1cm},
            every node/.style={draw, circle, minimum size=0, align=center}
        ]
        \node[fill=black] {}
            child {
            		node[draw=none, minimum size=0] {$\bar{n}$}
            }
            child {
            		node[fill=black] {}
            		child {
            			node[fill=black] {}
					child { node[draw=none, minimum size=0] {$\mathfrak{i}$} }
					child { node[draw=none, minimum size=0] {$\mathfrak{\tilde{i}}$} }
					edge from parent node[above=5pt, left=-6pt, draw=none] {$m_2$}
            		}
            		child {
            			node[fill=black] {}
            		    	child { node[draw=none, minimum size=0] {$v$} }
            		    	child { node[draw=none, minimum size=0] {$\tilde{v}$} }
            		    	edge from parent node[above=5pt, right=-6pt, draw=none] {$m_3$}
           		}
           		edge from parent node[above=5pt, right=-6pt, draw=none] {$m_1$}
           	};
\end{tikzpicture}
\caption{\justifying Tree structure used for the Holstein model.}
\label{fig:holstein_tree}
\end{figure}

For the HEOM+MCTDH calculation, a suitable tree structured must be devised. As discussed in \cref{subsec:heom+mctdh}, it is desirable to place any pair of physical and ancilla indices at the same bottom layer node of the tree. Furthermore, considering the coupling structure in the nanojunction setup, it is sensible to first contract the degrees of freedom within the system and then connect them to the $\mathbf{n}$-vector from the HEOM hierarchy structure. As a result, the tree structure depicted in \cref{fig:holstein_tree} is employed. The consolidated state vector is thus expanded as
\begin{align}
\ket{\Psi(t)}&=\sum_{\bar{n}}\sum_{k_1=1}^{m_1}A^1_{\bar{n},k_1}(t)\ket{\bar{n}}\otimes\ket{\phi^{1;1}_{k_1}(t)}, \\
\ket{\phi^{1;1}_{k_1}(t)}&=\sum_{k_2=1}^{m_2}\sum_{k_3=4}^{m_3}A^{2;1}_{k_1;k_2,k_3}(t)\ket{\phi^{2;1,1}_{k_2}(t)}\otimes\ket{\phi^{2;1,2}_{k_3}(t)}, \\
\ket{\phi^{2;1,1}_{k_2}(t)}&=\sum_{\mathfrak{i}\in\{0,1\}}\sum_{\mathfrak{\tilde{i}}\in\{0,1\}}A^{3;1,1}_{k_2;\mathfrak{i},\mathfrak{\tilde{i}}}(t)\ket{\mathfrak{i}}\otimes\ket{\mathfrak{\tilde{i}}}, \\
\ket{\phi^{2;1,2}_{k_3}(t)}&=\sum_{v=0}^{N_\mathrm{vib}-1}\sum_{\tilde{v}=0}^{N_\mathrm{vib}-1}A^{3;1,2}_{k_3;v,\tilde{v}}(t)\ket{v}\otimes\ket{\tilde{v}},
\end{align} where $\ket{\mathfrak{i}}$, $\ket{\mathfrak{\tilde{i}}}$ are the occupied (unoccupied) electronic state for $\mathfrak{i}=1$ ($\mathfrak{i}=0$), and $\ket{v}$, $\ket{\tilde{v}}$ are the vibrational states with $v$ quanta. The index $\bar{n}$ is a combination of the $m$ indices in the $\mathbf{n}$-vector. All calculations for the Holstein model are conducted using a vibrational basis of $N_\mathrm{vib}=40$ basis states. If not indicated otherwise, six Padé poles are used for the decomposition of the bath correlation functions. For $\Gamma_K=10\,\meV$ ($100\,\meV$), the calculations converged at the second (third) tier of the HEOM hierarchy.

The SPF basis sizes $m_1$, $m_2$ and $m_3$ are determined via the SPF basis size optimization procedure described in \cref{subsec:heom+mctdh}. Since $m_2$ has a maximum basis size of only $4$, we keep it fixed at this value and apply the optimization only to $m_1$ and $m_3$. For these, we choose initial SPF basis sizes of $m_{1/3}^{(0)}=5$. We use a checking interval of $t_\mathrm{check}=50\,\mathrm{a.u.}\approx1.2\,\mathrm{fs}$, basis increments of $m_{1/3}^{(+)}=2$, and recalculate $N_\mathrm{redo}=5$ checking intervals after any basis extension. This optimization procedure is repeated for different values of the natural population threshold, $10^{-4}\geq\delta\geq10^{-6}$, to assess how the threshold affects the results. For better comparability, the accuracy parameters of the CMF integrator are kept fixed at $\epsilon=10^{-6}$ and $\epsilon_\mathrm{CMF}=10^{-4}$.

\begin{figure*}
\begin{figure}[H]
\begin{minipage}{\textwidth}
\centering
\includegraphics[scale=1]{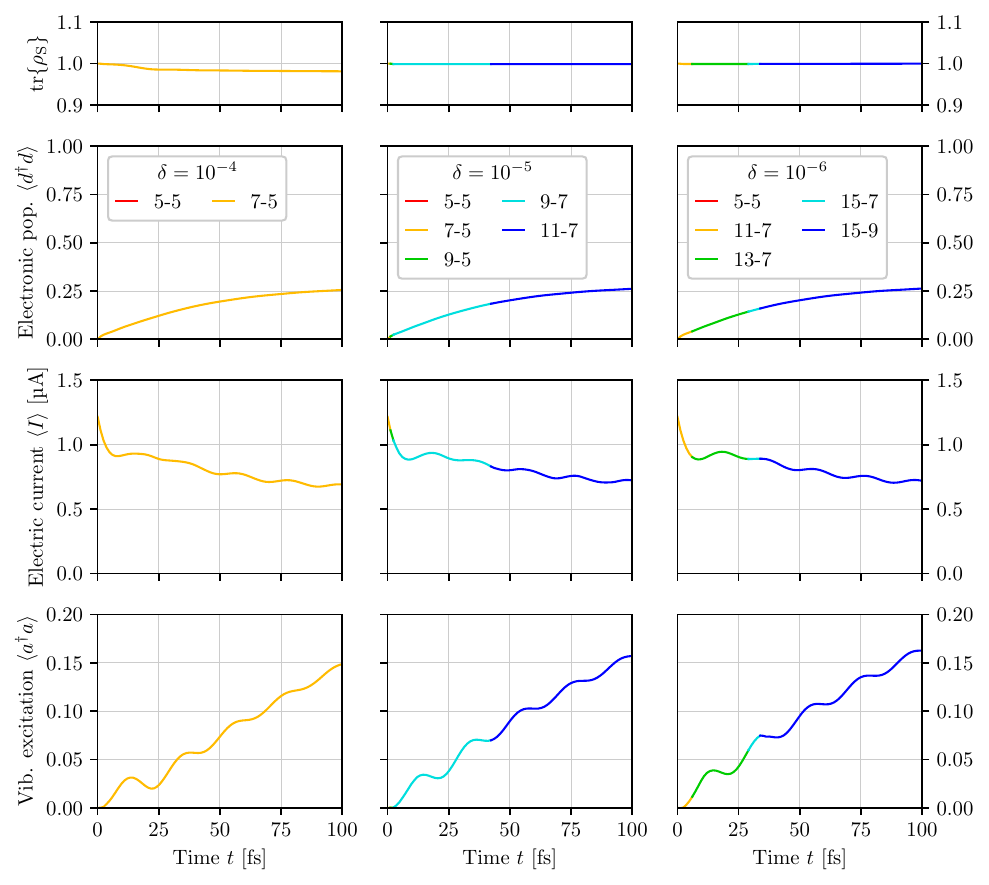}
\caption{\justifying Short-term dynamics of the Holstein model 1 for $\Gamma_K=10\,\meV$ and $\Phi=0.5\,\mathrm{V}$. From left to right, the columns depict simulations with various thresholds $\delta$ for the natural populations of the SPFs. A color change within a column represents an extension of the SPF basis sizes. The legends indicate the respective SPF basis sizes for a given color: The first number is $m_1$, the second is $m_3$; $m_2$ is fixed at $4$. From top to bottom, the rows depict the trace of the reduced density operator, $\tr\{\rhosys\}$, the electronic population, $\langle d^\dagger d\rangle$, the electric current, $\langle I\rangle$, and the vibrational excitation, $\langle a^\dagger a\rangle$.}
\label{fig:holstein_SPF}
\end{minipage}
\end{figure}
\end{figure*}

\Cref{fig:holstein_SPF} depicts the results of this process for Holstein model 1 with $\Gamma_K=10\,\meV$ and $\Phi=0.5\,\mathrm{V}$ for the first $100\,\mathrm{fs}$ of propagation time. It can be seen that for a large threshold of $\delta=10^{-4}$, only one basis extension is carried out during the $100\,\mathrm{fs}$ of propagation. Consequently, the simulation accuracy is limited, which manifests itself as $\tr\{\rhosys\}\neq1$ at later propagation times. For smaller thresholds, more basis extensions occur---up to four extensions within $100\,\mathrm{fs}$ for $\delta\leq10^{-5}$. For $\delta=10^{-6}$, SPF basis sizes of $m_1=15$ and $m_3=9$ are reached to ensure the threshold is met. The calculation accuracy is improved for the smaller $\delta$, as illustrated by the fact that $\tr\{\rhosys\}\approx1$ for the entire propagation time. Furthermore, there is barely a visible difference between the observable dynamics for $\delta=10^{-5}$ and $\delta=10^{-6}$, indicating that convergence has been reached. The results for $\delta=10^{-6}$ also demonstrate the recalculation of time steps, which leads to the basis sizes increasing from $m_1=5$, $m_3=5$ right to $m_1=11$, $m_3=7$. Given that we use increments of $m_{1/3}^{(+)}=2$, this can only occur due to the recalculation of time steps.

\begin{figure}
\centering
\includegraphics[scale=1]{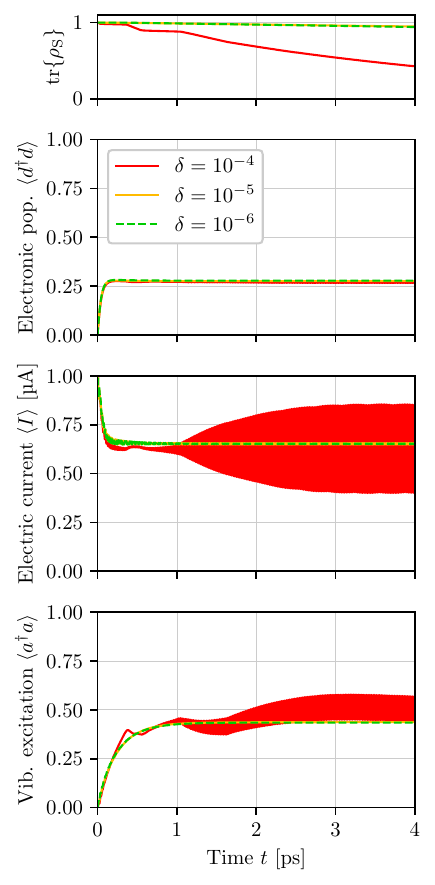}
\caption{\justifying Long-term dynamics of the Holstein model 1 for $\Gamma_K=10\,\meV$ and $\Phi=0.5\,\mathrm{V}$. From top to bottom, the rows depict the trace of the reduced density operator, $\tr\{\rhosys\}$, the electronic population, $\langle d^\dagger d\rangle$, the electric current, $\langle I\rangle$, and the vibrational excitation, $\langle a^\dagger a\rangle$. The large red areas are the result of rapid oscillations in the observables.}
\label{fig:holstein_convergence}
\end{figure}

In the next step, we consider longer propagation times, in order to allow the dynamics to evolve to the steady state. \Cref{fig:holstein_convergence} depicts $4\,\mathrm{ps}$ of propagation time with the same optimization procedure. It can be seen that for $\delta\leq10^{-5}$, the observables indeed converge and $\tr\{\rhosys\}$ stays close to $1$ even over long propagation times. For the larger $\delta$, however, the loss in accuracy leads to unphysical oscillations in the observables and a significant decrease of $\tr\{\rhosys\}$. Considering the findings in \cref{fig:holstein_convergence}, we choose $\delta=10^{-6}$ (or smaller) in the following.

\begin{figure}
\centering
\includegraphics[scale=1]{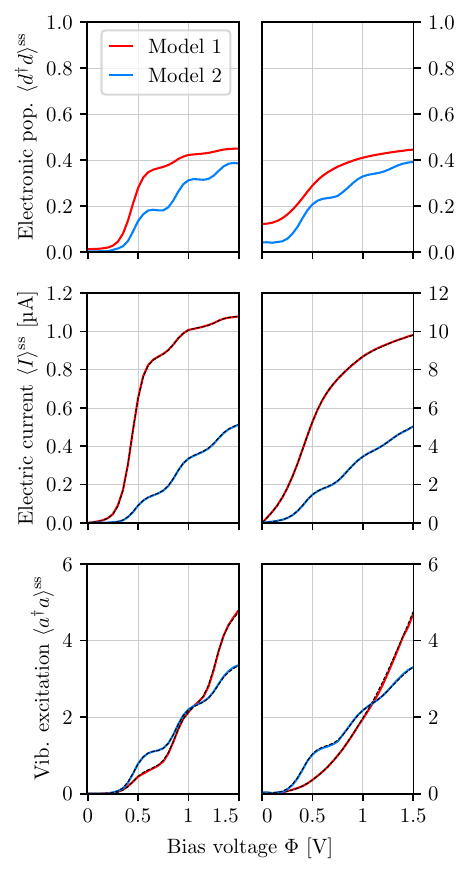}
\caption{\justifying Steady-state transport properties of the Holstein model 1 (red) and model 2 (blue) as a function of bias voltage $\Phi$. The left column depicts $\Gamma_K=10\,\meV$, the right column depicts $\Gamma_K=100\,\meV$. From top to bottom, the rows depict the mean electronic population, $\langle d^\dagger d\rangle^\mathrm{ss}$, mean electric current, $\langle I\rangle^\mathrm{ss}$, and mean vibrational excitation, $\langle a^\dagger a\rangle^\mathrm{ss}$. The black dashed lines depict a comparison calculation performed with an iterative steady-state solver.\cite{kaspar_efficient_2021}}
\label{fig:holstein_ss}
\end{figure}

We continue by calculating the steady-state transport properties for the Holstein model 1 and model 2 for a range of bias voltages, $\Phi=0.0-1.5\,\mathrm{V}$. Details of the SPF basis size optimization for these steady-state calculations are provided in \cref{appendix:spf_holstein}. The resulting steady-state values are shown in red (model 1) and blue (model 2) in \cref{fig:holstein_ss}. It can be seen that the results agree well with a comparison calculation (dashed black lines) performed with an iterative steady-state solver.\cite{kaspar_efficient_2021} In particular, the HEOM+MCTDH calculation correctly captures the characteristic step structure of the observables, which arises when the chemical potential of the left electrode exceeds the effective vibronic levels, $\varepsilon_1+n\cdot\omega$, thereby opening additional transport channels involving excitation of $n$ vibrational quanta in the nanojunction. It also reproduces the Franck-Condon blockade,\cite{koch_franck-condon_2005} i.e., the suppression of electric current at stronger vibronic coupling, as reflected in the lower current of model 2 (blue) compared to model 1 (red). Finally, for the larger electrode coupling, $\Gamma_K=100\,\meV$, the step structure is broadened due to stronger molecule-electrode hybridization.

This benchmark demonstrates the accuracy of the HEOM+MCTDH approach and provides the basis for its application to a more complex system in the following section.

\FloatBarrier

\subsection{Vibronic coupling nanojunction model}\label{subsec:2L2M}

In this section, we demonstrate the capabilities of the HEOM+MCTDH approach by considering a more complex molecular nanojunction model with strong molecule-electrode coupling comprising two electronic sites and two vibrational modes. Specifically, we consider the model studied in Refs.~\onlinecite{lu_blowing_2010,mack_vibrational_2025}, for which mixed quantum-classical methods have predicted a vibrational instability. Until now, there has been no fully quantum confirmation of this vibrational instability.

The model comprises two electronic sites that act as a chain junction, i.e., only site $1$ is coupled to the left electrode, and only site $2$ is coupled to the right electrode. Consequently, we have $V_\mathrm{L}^{(1)}=V_\mathrm{R}^{(2)}=\sqrt{\Gamma_K/2\pi}$ and $V_\mathrm{L}^{(2)}=V_\mathrm{R}^{(1)}=0$. The coupling between the electronic sites depends linearly on the displacement of the first vibrational mode, represented by $x_1=(a_1^\dagger+a_1)/\sqrt{2}$. The energy difference between the sites is modulated linearly by the displacement of the second mode, $x_2=(a_2^\dagger+a_2)/\sqrt{2}$. The system Hamiltonian, $\Hsys$, is given by
\begin{align}
\Hsys&=\varepsilon_1(x_2)d_1^\dagger d_1+\varepsilon_2(x_2)d_2^\dagger d_2+\tau(x_1)\left(d_1^\dagger d_2+d_2^\dagger d_1\right) \nonumber \\
&\quad+\omega_1a_1^\dagger a_1+\omega_2a_2^\dagger a_2,
\end{align} with
\begin{align}
\tau(x_1)&=\tau_0+\sqrt{2}\lambda_1x_1, \\
\varepsilon_1(x_2)&=\varepsilon_0+\sqrt{2}\lambda_2x_2=-\varepsilon_2(x_2), \\
\omega_1&=\omega_0+\frac{1}{2}\Delta\omega,\quad\omega_2=\omega_0-\frac{1}{2}\Delta\omega.
\end{align}

This model has been explored extensively using mixed quantum-classical methods,\cite{lu_blowing_2010,mack_vibrational_2025} where the electronic degrees of freedom are treated quantum-mechanically while the vibrational degrees of freedom are treated classically. In Ref.~\onlinecite{mack_vibrational_2025}, for example, approaches combining NEGFs for the electronic degrees of freedom and electronic friction with Langevin dynamics (LD) for the vibrations were used to examine the steady-state vibrational excitation as the detuning, $\Delta\omega=\omega_1-\omega_2$, of the unperturbed vibrational eigenenergies approaches zero. These investigations showed that, above a critical voltage, the limit $\Delta\omega\rightarrow0$ introduces vibrational instability, such that the vibrational excitation of both modes grows rapidly beyond the limit of a stable molecular nanojunction. However, although this effect has been well explored using such approximate methods where the large vibrational excitation was attributed to the influence of non-conservative current-induced electronic forces, no fully quantum-mechanical calculation of this model has been reported so far, primarily due to the large vibrational basis sizes required to model the regimes of instability and the large molecule-electrode coupling of $\Gamma_K=1\,000\,\meV$ considered in Refs.~\onlinecite{lu_blowing_2010,mack_vibrational_2025}.

The present HEOM+MCTDH approach can provide the first fully quantum-mechanical study of this nanojunction model, since the multi-layer MCTDH propagation allows for the use of large primitive bases, which are necessary to model the expected vibrational instability. In order to make this system more numerically accessible, we nevertheless consider a slightly modified parameter set in the following. Our parameters are given in \cref{tab:2L2M_parameters}. To connect with the previous mixed quantum-classical work on this model, we will also perform a comparison calculation for the new parameter set based on the NEGF-LD method outlined in Ref.~\onlinecite{mack_vibrational_2025}.

\begin{table}[H]
\centering
\begin{tabular}{|c|r|}
\hline
Parameter & \multicolumn{1}{|c|}{Value [meV]} \\
\hline
$\varepsilon_0$ & $0\hphantom{.00}$ \\
$\tau_0$ & $-200\hphantom{.00}$ \\
$\omega_0$ & $20\hphantom{.00}$ \\
$\Delta\omega$ & $0-3\hphantom{.00}$ \\
$\lambda_1$ & $40\hphantom{.00}$ \\
$\lambda_2$ & $40\hphantom{.00}$ \\
$\kB T$ & $25.8\hphantom{0}$ \\
$\Gamma_K$ & $500\hphantom{.00}$ \\
\hline
\end{tabular}
\caption{\justifying Parameters used for the vibronic coupling nanojunction model.}
\label{tab:2L2M_parameters}
\end{table}

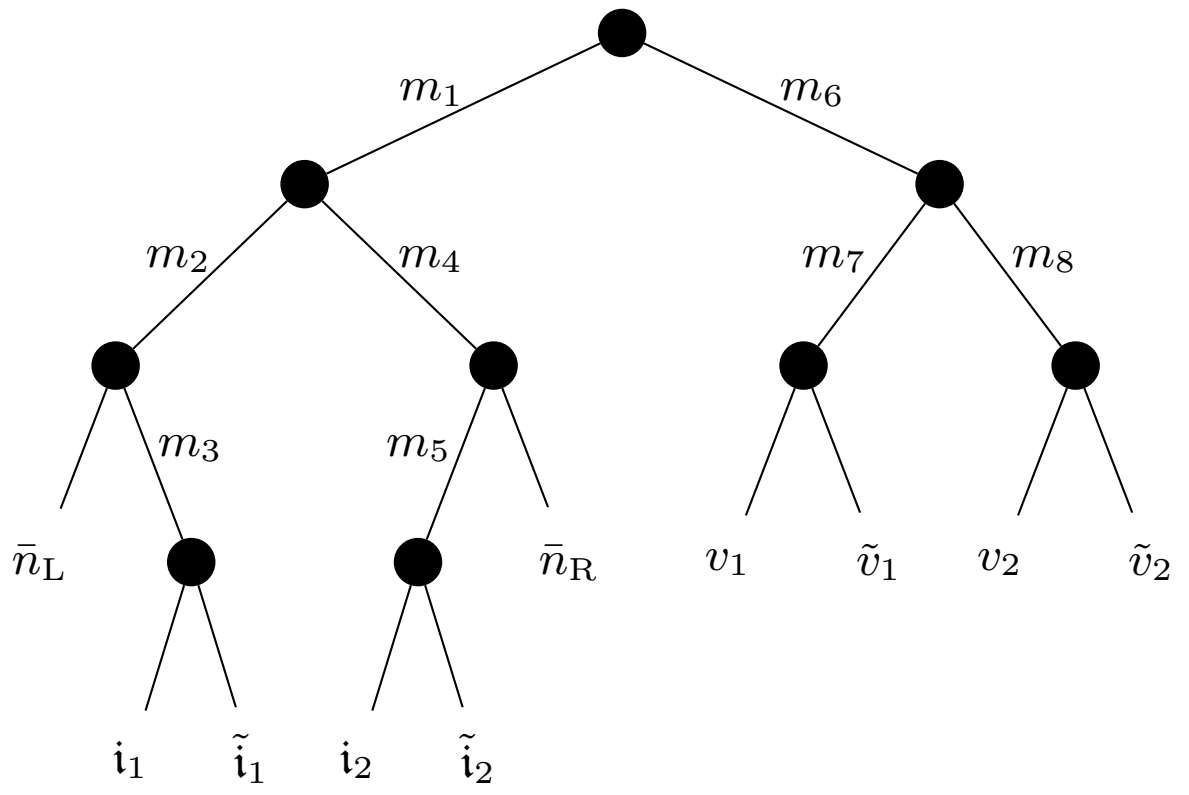
\begin{figure}
\centering
\begin{tikzpicture}
        [
        	grow=down,
            level 1/.style={sibling distance=4.2cm, level distance=1cm},
            level 2/.style={sibling distance=2.5cm, level distance=1.2cm},
            level 3/.style={sibling distance=1cm, level distance=1.3cm},
            level 4/.style={sibling distance=.8cm, level distance=1.3cm},
            every node/.style={draw, circle, minimum size=0, align=center}
        ]
         \node[fill=black] {}
         	child { node[fill=black] {}
				child { node[fill=black] {}
					child { node[draw=none, minimum size=0] {$\bar{n}_\mathrm{L}$} }
            			child { node[fill=black] {}
							child { node[draw=none, minimum size=0] {$\mathfrak{i}_1$} }
							child { node[draw=none, minimum size=0] {$\mathfrak{\tilde{i}}_1$} }
							edge from parent node[above=3pt, right=-4pt, draw=none] {$m_3$}
                		}
                		edge from parent node[above=3pt, left=-5pt, draw=none] {$m_2$}
            		}
            		child { node[fill=black] {}
            			child { node[fill=black] {}
							child { node[draw=none, minimum size=0] {$\mathfrak{i}_2$} }
							child { node[draw=none, minimum size=0] {$\mathfrak{\tilde{i}}_2$} }
							edge from parent node[above=3pt, left=-4pt, draw=none] {$m_5$}
                		}
					child { node[draw=none, minimum size=0] {$\bar{n}_\mathrm{R}$} }
                		edge from parent node[above=3pt, right=-5pt, draw=none] {$m_4$}
            		}
            		edge from parent node[above=3pt, left=-5pt, draw=none] {$m_1$}
            	}   
           	child { node[fill=black] {}
            		child[sibling distance=1.8cm] { node[fill=black] {}
					child { node[draw=none, minimum size=0] {$v_1$} }
					child { node[draw=none, minimum size=0] {$\tilde{v}_1$} }
					edge from parent node[above=3pt, left=-4pt, draw=none] {$m_7$}
                }
				child[sibling distance=1.8cm] { node[fill=black] {}
					child { node[draw=none, minimum size=0] {$v_2$} }
					child { node[draw=none, minimum size=0] {$\tilde{v}_2$} }
					edge from parent node[above=3pt, right=-4pt, draw=none] {$m_8$}
                }
                edge from parent node[above=3pt, right=-5pt, draw=none] {$m_6$}
            };
\end{tikzpicture}
\caption{\justifying Tree structure used for the vibronic coupling nanojunction model with split left/right bath indices, $\bar{n}_\mathrm{L}$ and $\bar{n}_\mathrm{R}$. Note that by design of the tree structure $m_1=m_6$.}
\label{fig:2L2M_tree}
\end{figure}

In the extended twin space, we obtain again \cref{eq:heom_condensed_wbl} as the equation of motion for the extended state vector, $\ket{\Psi(t)}$, and the super-Hamiltonian $\mathscr{H}_\mathrm{WBL}$ as given in \cref{eq:super_hamiltonian,eq:super_hamiltonian_wbl}. For the HEOM+MCTDH calculation, an appropriate tree structure must be devised. Since we are dealing with a chain-type model, it is advantageous to partition the $\mathbf{n}$-vector into a separate left and right electrode index, $\bar{n}_\mathrm{L}$ and $\bar{n}_\mathrm{R}$, and to assign individual indices to the electronic sites in the system. As such, $\mathfrak{i}_j$ and $\mathfrak{\tilde{i}}_j$ are the electronic indices of the system and $\mathfrak{i}_j=1$ ($\mathfrak{i}_j=0$) denotes the occupied (unoccupied) electronic state of the $j$th site. In order to exploit the separate indices in the MCTDH calculation, $\bar{n}_\mathrm{L}$ and $\mathfrak{i}_1$, respectively $\bar{n}_\mathrm{R}$ and $\mathfrak{i}_2$, are first contracted. Next, the electronic indices are contracted together and finally linked to the vibrational degrees of freedom, where $v_\alpha$ and $\tilde{v}_\alpha$ are the vibrational indices and denote the vibrational eigenstate of mode $\alpha$ with $v$ quanta. The resulting tree structure is depicted in \cref{fig:2L2M_tree}. Since this model requires propagation times of several tens of picoseconds, the HEOM hierarchy is truncated at the third tier.

\begin{figure}
\centering
\includegraphics[scale=1]{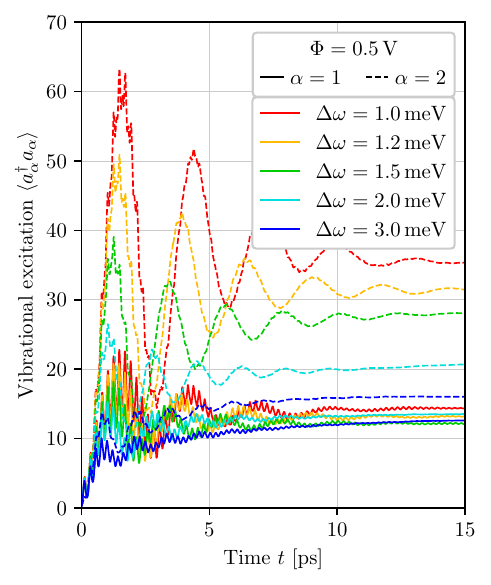}
\caption{\justifying Transient dynamics of the average vibrational excitations, $\langle a_\alpha^\dagger a_\alpha\rangle$, of the vibronic coupling nanojunction model for $\Gamma_K=500\,\meV$ and $\Phi=0.5\,\mathrm{V}$. The solid (dashed) lines depict the excitation of the first (second) mode.}
\label{fig:2L2M_excs}
\end{figure}

\begin{figure}
\centering
\includegraphics[scale=1]{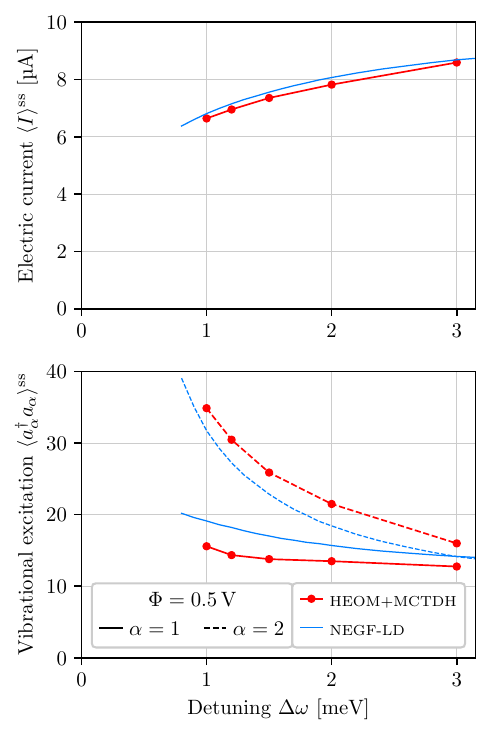}
\caption{\justifying Steady-state electric current, $\langle I\rangle^\mathrm{ss}$, and vibrational excitations, $\langle a_\alpha^\dagger a_\alpha\rangle^\mathrm{ss}$, of the vibronic coupling nanojunction model for $\Gamma_K=500\,\meV$ and $\Phi=0.5\,\mathrm{V}$. The connected red dots represent fully quantum calculations using the present HEOM+MCTDH approach. The blue lines show comparison calculations based on the mixed quantum-classical NEGF-LD approach.\cite{mack_vibrational_2025} In the bottom panel, the solid (dashed) lines depict the steady-state excitation of the first (second) mode.}
\label{fig:2L2M_ss}
\end{figure}

\Cref{fig:2L2M_excs} shows the first $15\,\mathrm{ps}$ of the resulting transient dynamics of the average vibrational excitations for several different detunings, $\Delta\omega=1-3\,\meV$, at a fixed bias voltage of $\Phi=0.5\,\mathrm{V}$. Details of the SPF basis size optimization used in these calculations are provided in \cref{appendix:spf_2L2M}. The dynamics differ markedly between large and small vibrational detuning. For the largest detuning considered, $\Delta\omega=3\,\meV$, the vibrational excitations exhibit a gradual increase toward their steady-state values. The increase is superimposed with low-amplitude high-frequency oscillations with a period of about $h/\omega_{1/2}\approx0.2\,\mathrm{ps}$, corresponding to the natural frequencies of the two vibrational modes, and a medium-amplitude low-frequency oscillation with period of about $h/\Delta\omega\approx1.4\,\mathrm{ps}$. The high-frequency oscillation is more pronounced in the first mode.

For smaller detunings, the transient dynamics change qualitatively. The vibrational excitations initially rise sharply within the first $1-2\,\mathrm{ps}$, reaching values far above the steady-state excitations obtained at later times. This initial growth is followed by an oscillatory decay toward the steady state, dominated by a high-amplitude low-frequency oscillation with a period of approximately $h/\Delta\omega$. For example, for $\Delta\omega=1\,\meV$, this corresponds to a period of $h/1\,\meV\approx4\,\mathrm{ps}$. At the smallest detuning considered, the second mode reaches a transient mean excitation of more than $60$, whereas its steady-state mean excitation is only about $35$.

Overall, decreasing the detuning, $\Delta\omega\rightarrow0$, leads to a strong increase in vibrational excitation, most prominently in the second mode. Moreover for small detunings, $\Delta\omega\leq2\,\meV$, the high-amplitude low-frequency beat appears to govern the timescale of the oscillatory decay. This results in very long relaxation times of more than $20\,\mathrm{ps}$ for reaching the steady state at $\Delta\omega\leq2\,\meV$, and suggests a further increase in relaxation times as $\Delta\omega\rightarrow0$, possibly indicating the eventual loss of a steady state at $\Delta\omega=0$.

The resulting steady-state vibrational excitations and electric current are depicted in \cref{fig:2L2M_ss}, together with comparison results obtained using the mixed quantum-classical NEGF-LD approach.\cite{mack_vibrational_2025} Both the fully quantum HEOM+MCTDH calculation and the mixed quantum-classical NEGF-LD calculation predict similar vibrational excitations and an almost identical electric current. In particular, both methods reproduce the strong increase in vibrational excitation as the two modes approach degeneracy, $\Delta\omega\rightarrow0$, indicating vibrational instability.

The HEOM+MCTDH results provide the first fully quantum confirmation of this vibrational instability. At the same time, the close agreement with the NEGF-LD results supports the validity and accuracy of the mixed quantum-classical description for the nanojunction model considered here. Nevertheless, minor differences between the two approaches remain. In particular, the HEOM+MCTDH calculation predicts consistently lower excitations for the first vibrational mode and higher excitations for the second mode.

Finally, we note that the electric current through the nanojunction decreases with decreasing detuning, $\Delta\omega$; in other words, it behaves inversely to the vibrational excitations. This behavior is reminiscent of vibrationally induced negative differential resistance, where increasing the bias voltage can simultaneously enhance vibrational excitation while suppressing the electric current.\cite{hartle_resonant_2011} Here, however, the effect occurs at a fixed bias voltage as a result of varying the detuning of the vibrational eigenenergies.

\section{Conclusion}\label{sec:conclusion}

In this work, we have presented a novel approach for propagating the hierarchical equations of motion (HEOM) for fermionic environments. Starting from the existing twin-space formulation of the HEOM,\cite{borrelli_expanding_2021,ke_hierarchical_2022} we described the mapping that enables the fermionic HEOM to be propagated using the (multi-layer) multi-configurational time-dependent Hartree method (MCTDH).\cite{meyer_multi-configurational_1990,manthe_wavepacket_1992,wang_multilayer_2003,manthe_multilayer_2008} We further provided implementation details for the resulting HEOM+MCTDH approach. In particular, we described the SPF basis size optimization procedure employed and discussed constraints imposed on the design of the multi-layer MCTDH tree structures. We then validated the accuracy of the resulting HEOM+MCTDH approach by considering a single-site Holstein-type molecular nanojunction model for which reference data are available.\cite{schinabeck_hierarchical_2016} Finally, we demonstrated the performance of the HEOM+MCTDH approach by presenting fully quantum calculations for a molecular nanojunction model with strong molecule-electrode coupling comprising two electronic sites and two vibrational modes.\cite{lu_blowing_2010,mack_vibrational_2025} For this nanojunction model, the HEOM+MCTDH calculation provides the first fully quantum confirmation of the existence of a vibrational instability near the degenerate case, $\Delta\omega\rightarrow0$, which had previously been predicted using mixed quantum-classical methods.\cite{lu_blowing_2010,mack_vibrational_2025}

Overall, these results establish the HEOM+MCTDH approach as a powerful method for the propagation of the fermionic HEOM. Future work will focus in particular on exploring the potential of HEOM+MCTDH for simulating nanojunction models of greater complexity. Applications of HEOM+MCTDH to scattering processes\cite{preston_nonadiabatic_2025} or catalytic reactions on metallic surfaces are also conceivable. Moreover, an extension to purely bosonic environments or hybrid fermionic-bosonic environments may be possible in analogy to MPS-based approaches to HEOM.\cite{ke_hierarchical_2022}

\section*{Acknowledgements}

This work was supported by the German Research Foundation (DFG) through Research Unit FOR5099, and from the state of Baden-Württemberg through bwHPC and the DFG through Grant No.~INST 40/575-1 FUGG (Justus 2 cluster). R.J.P. acknowledges support from the Alexander von Humboldt Foundation.

\appendix

\section{SPF basis size optimization for the Holstein model}\label{appendix:spf_holstein}

In the following, we list the details of the SPF basis size optimization as used for the calculation of the steady-state transport properties of the Holstein model 1 and model 2 depicted in \cref{fig:holstein_ss}.

\begin{table}
\centering
\begin{tabular}{|c|c|c|c|c|}
\hline
\multirow{2}{*}{Voltage range} & \multicolumn{3}{|c|}{Accuracy parameters} & \multirow{2}{*}{Padé poles} \\
 & $\epsilon$ & $\epsilon_\mathrm{CMF}$ & $\delta$ & \\
\hline
\multicolumn{5}{|c|}{Model 1 at $\Gamma_K=10\,\meV$} \\
\hline
$1.5\,\mathrm{V}\geq\Phi\geq1.0\,\mathrm{V}$ & $\ 10^{-6}\ $ & $\ 10^{-4}\ $ & $\ 10^{-6}\ $ & 8 poles \\
$1.0\,\mathrm{V}>\Phi>0.2\,\mathrm{V}$ & $10^{-6}$ & $10^{-4}$ & $10^{-6}$ & 6 poles \\
$0.2\,\mathrm{V}\geq\Phi\geq0.0\,\mathrm{V}$ & $10^{-7}$ & $10^{-5}$ & $10^{-7}$ & 6 poles \\
\hline
\multicolumn{5}{|c|}{Model 1 at $\Gamma_K=100\,\meV$} \\
\hline
$1.5\,\mathrm{V}\geq\Phi\geq1.0\,\mathrm{V}$ & $10^{-6}$ & $10^{-4}$ & $10^{-6}$ & 8 poles \\
$1.0\,\mathrm{V}>\Phi\geq0.0\,\mathrm{V}$ & $10^{-6}$ & $10^{-4}$ & $10^{-6}$ & 6 poles \\
\hline
\multicolumn{5}{|c|}{Model 2 at $\Gamma_K=10\,\meV$} \\
\hline
$1.5\,\mathrm{V}\geq\Phi\geq0.0\,\mathrm{V}$ & $10^{-7}$ & $10^{-5}$ & $10^{-7}$ & 8 poles \\
\hline
\multicolumn{5}{|c|}{Model 2 at $\Gamma_K=100\,\meV$} \\
\hline
$1.5\,\mathrm{V}\geq\Phi\geq0.0\,\mathrm{V}$ & $10^{-6}$ & $10^{-4}$ & $10^{-6}$ & 8 poles \\
\hline
\end{tabular}
\caption{\justifying Accuracy parameters and number of Padé poles used for calculating the steady-state transport properties of Holstein model 1 and model 2.}
\label{tab:holstein_ss_parameters}
\end{table}

The optimization procedure from \cref{subsec:heom+mctdh} is applied to determine the basis sizes $m_1$ and $m_3$. The basis size $m_2$ is kept at its maximum value of $4$ like in the previous calculations for the Holstein model. For efficiency reasons, we use initial basis sizes of $m_{1/3}^{(0)}=10$, and increments of $m_{1/3}^{(+)}=5$ for the steady-state calculations. For large voltages $\Phi\geq1\,\mathrm{V}$, more than six Padé poles are required for the decomposition of the bath correlation function to ensure the Fermi function is accurately represented within the bias window. For model 2, this is required already for smaller voltages because of the large electronic on-site energy, $\varepsilon_0=1050\,\meV$. Furthermore, for $\Gamma_K=10\,\meV$, some voltage ranges require smaller integration accuracies of $\epsilon=10^{-7}$ and $\epsilon_\mathrm{CMF}=10^{-5}$ and a correspondingly smaller natural population threshold of $\delta=10^{-7}$ to achieve converged steady-state results. The exact parameter combinations are listed in \cref{tab:holstein_ss_parameters}.

\section{SPF basis size optimization for the vibronic coupling nanojunction model}\label{appendix:spf_2L2M}

In the following, we discuss the details of the SPF basis size optimization as used for the vibronic coupling nanojunction model.

All calculations for this model are conducted using seven Padé poles for the decomposition of the bath correlation function, which means there are $2\cdot7=14$ virtual fermionic modes for each combined index, $\bar{n}_\mathrm{L}$ and $\bar{n}_\mathrm{R}$. To determine the SPF basis sizes $m_1,\dots,m_8$, we apply the optimization procedure from \cref{subsec:heom+mctdh}. Note that by design of the tree structure $m_1=m_6$. For $m_3$ and $m_5$, the maximum value is only $4$. Thus, we keep them fixed at this maximum value. We initialize the remaining SPF bases with $m_j^{(0)}=10$ basis states, use a checking interval of $t_\mathrm{check}=50\,\mathrm{a.u.}\approx1.2\,\mathrm{fs}$, basis increments of $m_j^{(+)}=5$, and recalculate $N_\mathrm{redo}=5$ checking intervals after any basis extension.

Employing these settings, we propagate the model to the steady state for several different detunings, $\Delta\omega=1-3\,\meV$, at a fixed bias voltage of $\Phi=0.5\,\mathrm{V}$. The required vibrational basis size increases with decreasing detuning, from $N_\mathrm{vib}=60$ at $\Delta\omega=3\,\meV$ to $N_\mathrm{vib}=120$ at $\Delta\omega=1\,\meV$. Smaller detunings are not considered, since the required vibrational basis size increases rapidly with decreasing $\Delta\omega$, making fully quantum calculations prohibitively expensive and likely unfeasible for $\Delta\omega\leq0.5\,\meV$.

Since the model requires propagation times to the steady state of more than $20\,\mathrm{ps}$, the HEOM hierarchy is truncated at the third tier for all detunings. For $\Delta\omega\geq2\,\meV$, we set the threshold for the natural populations of the SPFs to $\delta=10^{-7}$ and the accuracy parameters of the CMF integrator to $\epsilon=10^{-7}$ and $\epsilon_\mathrm{CMF}=10^{-5}$. For smaller detunings, $\Delta\omega\leq1.5\,\meV$, the high vibrational excitation reached requires a compromise between accuracy and computational cost. In these cases, it proved beneficial to employ looser accuracy parameters of $\delta=10^{-6}$, $\epsilon=10^{-6}$, and $\epsilon_\mathrm{CMF}=10^{-4}$ for an initial pre-relaxation of $10^6\,\mathrm{a.u.}\approx24.2\,\mathrm{ps}$, and apply the aforementioned stricter accuracy parameters only for the final relaxation of a few $\mathrm{ps}$.

\bibliographystyle{apsrev4-2}
\bibliography{paper.bib}

\end{document}